\documentclass[a4paper,12pt,twoside]{article} 
\usepackage{a4wide}
\usepackage{graphicx}
\usepackage{rotating}
\usepackage{amssymb}

\newcommand\jhep[3] {{{\it J. High Energy Phys.\ }{\bf #1} (#2) #3}}
\newcommand\npb[3] {{{\it Nucl.\ Phys.\ }{\bf B #1} (#2) #3}}
\newcommand\npps[3] {{{\it Nucl.\ Phys.\ }{\bf #1} {\it(Proc.\ Suppl.)} (#2) #3}}
\newcommand\plb[3] {{{\it Phys.\ Lett.\ }{\bf B #1} (#2) #3}}
\newcommand\prd[3] {{{\it Phys.\ Rev.\ }{\bf D #1} (#2) #3}}
\newcommand\prl[3] {{{\it Phys.\ Rev.\ Lett.\ }{\bf #1} (#2) #3}}
\newcommand\rmp[3] {{{\it Rev.\ Mod.\ Phys.\ }{\bf #1} (#2) #3}}
\newcommand\sjnp[3] {{{\it Sov.\ J.\ Nucl.\ Phys.\ }{\bf #1} (#2) #3}}

\begin{document}

{\flushright

FTUV-01-1114

}

\begin{center}
\noindent{\Large \tt \bf 
%------------------------------------------ Title --------------------

Observable Medium Effects For Atmospheric Neutrinos
 
%---------------------------------------------------------------------
}\vspace{10mm}
\renewcommand{\thefootnote}{\fnsymbol{footnote}}
\noindent{\large
%-------------------------------------- Author(s) --------------------
J. Bernab\'eu\footnote{Talk given at the EPS-Conference on HEPP, Budapest
(Hungary), July 2001, to be published in the Proceedings.} and
S. Palomares-Ruiz 
%---------------------------------------------------------------------
}\vspace{6mm}

\noindent{\it
%------------------------------------ Address(es) --------------------
  Departament de F\'{\i}sica Te\`orica 

  Universitat de Val\`encia  
      
  46100 Burjassot, Val\`encia, Spain
%---------------------------------------------------------------------
\\
}
\end{center}
\vspace{6mm}
\renewcommand{\thefootnote}{\arabic{footnote}}
\setcounter{footnote}{0}

\begin{abstract}
We discuss the possibility to observe matter effects in atmospheric
neutrino oscillations. The main conclusion is that an impact on the $\nu_{\mu}$
survival probability requires the action of the MSW resonance, which becomes
visible for baselines above $\sim$ 7000 km. The associated muon charge
asymmetry carries information on $\theta_{13}$ and the sign of $\Delta
m^{2}_{31}$.
\end{abstract}

\newpage

  \section{Introduction}

Present evidence for neutrino masses and mixings can be summarized as:
1) the atmospheric $|\Delta m^{2}_{31}| \sim (1-5) \cdot 10^{-3}
$eV$^2$ is associated with a mixing, $\theta_{23}$, near to maximal
\cite{SK}; 2) the solar $\Delta m^{2}_{21}$ prefers the LMA-MSW
solution \cite{solar}; CHOOZ reactor data \cite{chooz} give severe
limits for $|U_{e 3}|$. In this contribution we are going to discuss
that contrary to a wide spread belief, Earth effects on the propagation
of atmospheric neutrinos can become observable \cite{earth} even if
$|U_{e 3}|$ is small, but non-vanishing. This fact would allow to determine the
sign of $\Delta m^{2}_{31}$ \cite{barger}. For baselines $L$ smaller than the
Earth diameter, appropiate for atmospheric neutrinos, $\frac{\Delta
m^{2}_{21}}{4 E} L \equiv \Delta_{21} \ll 1$, so that we will neglect the
(1,2)-oscillating phase in vacuum against the (2,3)-one. This is a very good
aproximation, unless the high $\Delta m^{2}_{21}$-region of the LMA solution
turns out to be the solution to the solar problem. In that case we should take
into account corrections of order $O (\frac{\Delta m^{2}_{21}}{\Delta
m^{2}_{21}})$ (see eg. \cite{corrections}).  

In section 2 we discuss the correspondence between the determination of
the sign of $\Delta m^{2}_{31}$ and the observation of the Earth effects in a
transition involving $\nu_e$. The change expected in the neutrino spectrum and
mixing due to matter effects, both for $\sin{\theta_{13}} \equiv s_{13} = 0$
and $s_{13} \neq 0$, are pointed out. In the latter case, section 3 studies the
observability of the MSW-resonance, with a positive conclusion for baselines $L
\gtrsim$ 7000 km . Its impact on the survival probability, $\nu_{\mu}
\rightarrow \nu_{\mu}$, is pointed out. Section 4 gives an analysis of the
matter-induced CPT-odd asymmetry, together with the realistic charge-asymmetry
expected for atmospheric neutrinos. In section 5 we present some conclusions.

  \section{The neutrino spectrum in matter}

Current analyses leave us with two alternatives for the spectrum of the three
active neutrino species, either hierarchical or degenerate.  

The effective neutrino potential due to the charged current interaction of
$\nu_e$ with the electrons in the medium is \cite{wolf} $V \equiv \frac{a}{2 E}
= \sqrt{2} G_F N_e$, so that the effective hamiltonian, in the extreme
relativistic limit, is given by \cite{kuo-pan}

\begin{equation}
\label{h}
H = \frac{1}{2 E} \left\{ U \left( \begin{array}{ccc} 
0 & \; 0 & 0 \\ 0 & \; 0 & 0 \\ 0 & \; 0 & \Delta m^{2}_{31} \end{array}
\right) 
U^{\dagger} + \left( \begin{array}{ccc} 
a & 0 & 0 \\ 0 & 0 & 0 \\ 0 & 0 & 0 \end{array} \right) \right\}
\end{equation}

In going from $\nu$ to $\overline{\nu}$, there are matter-induced CP- and CPT-
odd effects associated with the change $a \rightarrow - a$. The additional
change U $\rightarrow$ U$^*$ is inoperative in the limit of (\ref{h}). The
effects we are going to discuss depend on the interference between the
different flavors and on the relative sign between $a$ and $\Delta
m^{2}_{31}$. As a consequence, an experimental distinction between the
propagation of $\nu$ and $\overline{\nu}$ (the sign of $a$) will determine the
sign of $\Delta m^{2}_{31}$. An appreciable interference will be present if and
only if there are appreciable matter effects. For atmospheric neutrinos, one
needs the ``connecting'' mixing $U_{e 3}$ between the $\nu_e$-flavor and the
$\nu_3$ mass eigenstate to show up.   

For $s_{13} = 0$, matter effects lead to a breaking of the (1,2)-degeneracy
such that $\tilde{\nu}_2$ coincides with $\nu_e$. The net effect is that
$\tilde{\nu}_1$ and $\tilde{\nu}_3$ in matter remain unaltered, i. e., as for 
vacuum (2,3), leading to the $\nu_{\mu} \rightarrow \nu_{\tau}$ indicated by
SK. The $\nu_e$-flavor decouples in matter, even if there was a large mixing in
the (1,2)-system, as shown by the solar experiments. No matter effects would
then be expected when starting with $\nu_{\mu}$. The CHOOZ limit \cite{chooz},
$\sin^2{2 \theta_{13}} \le 0.10$, is it then fatal?

For small $s_{13}$, even if the effects on the spectrum are expected to be
small, there could be a substantial mixing of $\nu_e$ with $\tilde{\nu}_3$ if
one is near to a situation of level-crossing. This would lead to a resonant MSW
behaviour \cite{msw}. 

\begin{equation}
\label{resonance}
\sin^2{2 \, \tilde{\theta}_{13}} = \frac{4 \, s^{2}_{13} \, c^{2}_{13}}{(\alpha
- \cos{2 \, \theta_{13}})^2 + 4 \, s^{2}_{13} \, c^{2}_{13}} \hspace{0.3cm},
\hspace{0.7cm} \alpha \equiv \frac{a}{\Delta m^{2}_{31}}
\end{equation}

But still $\langle \tilde{\nu}_1 | \nu_e \rangle = 0$, i.\ e., the $\nu_e$ has
no overlap with the lowest mass eigenstate in matter. This vanishing mixing in
matter is responsible for the absence of fundamental CP-violating effects, even
if there are three non-degenerate mass eigenstates in matter. In vacuum, the
absence of genuine CP-odd probabilities was due to the degeneracy $\Delta_{21}
= 0$. The step from vanishing $\Delta_{21}$ in vacuum to the vanishing mixing
$U_{e 1}$ in matter was termed a ``transmutation'' \cite{trans}.

For matter of constant density, there is no asymmetry associated with
time-reversal T, contrary to the matter-induced CP- and CPT- odd
asymmetries. As a consequence, the T-conjugated transitions ($\alpha \neq
\beta$) have equal appearance probabilities \cite{T}, $P(\nu_{\alpha}
\rightarrow \nu_{\beta}) = P(\nu_{\beta} \rightarrow \nu_{\alpha})$.

  \section{Observability of the MSW resonance}

Non-resonant important matter effects are induced in the non-leading appearance
channel governed by $\Delta_{31} \equiv \frac{\Delta m^{2}_{31}}{4 E} L$. The
corresponding probability for $\alpha \neq 1$ is given by \cite{earth}

\begin{equation}
\label{prob}
P(\nu_e \rightarrow \nu_{\mu}) \simeq \; s^{2}_{23} \; \frac{4 \,
s^{2}_{13}}{(1 - \alpha)^2} \; \sin^2{[\Delta_{31} \, (1 - \alpha)]}
\end{equation}

This probability has a good sensitivity to $s_{13}$. Besides the change in the
effective mixing, the interference pattern is modified to an oscillating phase,
$\tilde{\Delta}_{31} \simeq \Delta_{31} \, (1 - \alpha)$, which contains a
baseline dependence as a combination of the standard $L/E$, plus the (constant)
$\times \, L$ terms \cite{trans}. This is reminiscent of the Aharonov-Bohm
effect, able to see the potential in quantum-mechanical phases.

For a baseline of $L$ = 3000 km, appropiate for neutrino factories \cite{nuf},
there are non-resonant Earth-matter effects in the sub-leading appearance
probability. At the resonant energy $E_R \sim$ 9.6 GeV, there is no observable
trace of the resonance. This suppresion can be understood because a resonant
mixing is  associated with level-crossing, so that $\Delta \tilde{m}^{2}_{31}$
is minimum on the resonance and then, $\tilde{\Delta}_{31} \simeq 0$ for such
$L$.  

\begin{figure}[h]
\begin{center}
\includegraphics[width=10cm]{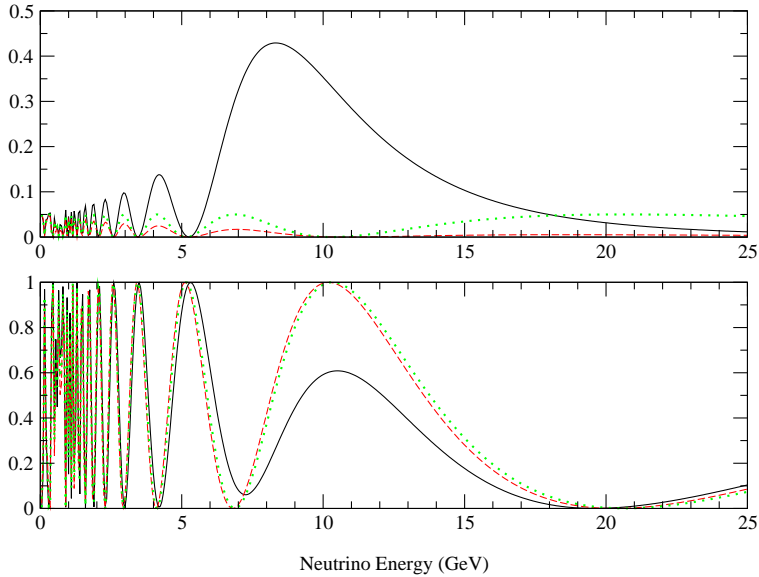} 
\caption{\label{L8000} Appearance (upper panel) and survival (lower panel)
probabilities, $P(\nu_{\mu} \rightarrow \nu_e)$ and $P(\nu_{\mu} \rightarrow
\nu_{\mu})$, for neutrinos (solid line), antineutrinos (dashed line) and vacuum
(dotted line), for $L = 8000$ km, $\Delta m^{2}_{31} = 3.2 \cdot 10^{-3}$
eV$^2$, $\sin^{2}{2 \theta_{23}} = 1$ and $\sin^{2}{2 \theta_{13}} = 0.1$.}
\end{center} 
\end{figure}

For atmospheric $\nu_{\mu}$ neutrinos, matter effects in the survival
probability $\nu_{\mu} \rightarrow \nu_{\mu}$ would be minute unless the
resonance shows up. The resonance was not apparent even at $L$ = 3000 km. Is
there a way out?

Again, a non-vanishing connecting mixing $s_{13} \neq 0$ provides the
solution. Along with it, there is a resonance width which, when discussed in
terms of the dimensionless parameter $\alpha$, is given by \vspace{-2mm}

\begin{equation}
\label{alpha}
\alpha_R = \cos{2 \, \theta_{13}} \hspace{0.3cm}, \hspace{0.7cm}
\Gamma_{\alpha} = 2 \; \sin{2 \, \theta_{13}}
\end{equation}

One discovers that the oscillating phase on the resonance is non-vanishing, but
given by the $L$-dependent relation

\begin{equation}
\label{res}
\tilde{\Delta}_{31 (R)} = \Delta_{31} \; \frac{\Gamma_{\alpha}}{2}
\end{equation}
  
If $L \ll L_{opt}$, with optimal $L$, $L_{opt}$, defined by $\tilde{\Delta}_{31
(R)} = \pi /2$, the resonance does not affect the oscillation probability. On
the contrary, around $L_{opt} = \frac{2 \; \pi}{\tilde{a} \; \tan{2 \,
\theta_{13}}}$, where $\tilde{a} = a/E$, the resonance becomes apparent and
$L_{opt}$ is independent of $\Delta m^{2}_{31}$, which determines the resonant 
energy. For $L = L_{opt}$, the maximum mixing is accompanied by maximum
oscillating factor.

Under these conditions, all channels would see the resonant effect. Contrary to
non-resonant matter effects, the resonance only affects the (anti)neutrino
channels if $\Delta m^{2}_{31} > 0 (< 0)$. For a  baseline $L$ = 8000 km, this
is shown in fig. \ref{L8000} for both $\nu_e \rightarrow \nu_{\mu}$ and
$\nu_{\mu} \rightarrow \nu_{\mu}$ channels. 

\begin{figure}[h]
\begin{center}
\includegraphics[width=10cm]{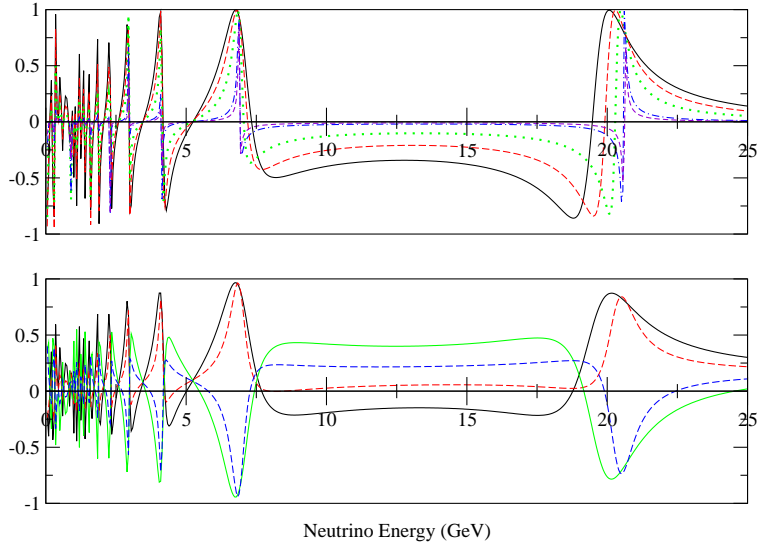}
\caption{\label{asym} Upper panel: CPT-asymmetry, $A_{CPT}$,
for different values of $\sin^{2}{2 \theta_{13}}$. From up to down: $\sin^{2}{2
\theta_{13}}$ = 0.005, 0.01, 0.05, 0.10, 0.16 and $\Delta m^{2}_{31} > 0$ as
all the plots are symmetric with respect to the horizontal axis when $\Delta
m^{2}_{31} < 0$. Lower panel: Charge-asymmetry, A, for $\sin^{2}{2
\theta_{13}}$ = 0.05 (dashed line) and 0.16 (solid line). The lower plots
correspond to $\Delta m^{2}_{31} > 0$ and the upper ones to $\Delta m^{2}_{31}
< 0$. For both panels, $L = 8000$ km, $\sin^{2}{2 \theta_{23}}$ = 1 and
$\left|\Delta m^{2}_{31}\right| = 3.2 \cdot 10^{-3}$ eV$^2$ . \label{asym}}  
\end{center}
\end{figure}

  \section{Charge asymmetries}

As discussed in section 3, matter effects distinguish neutrinos from
antineutrinos. It is convenient to present them in terms of CP-odd (for
appearance channels) and CPT-odd (for the survival probabilities)
asymmetries. In the limit $\Delta_{21} = 0$, there is no room for genuine CP
violation. The interaction with matter will generate an asymmetry effect,
however, which is not connected with the vacuum propagation. 

\noindent 
For $\nu_{\mu}$ and $\bar{\nu}_{\mu}$, one has 

\begin{equation}
\label{acpt}
A_{CPT} = \frac{P (\nu_{\mu} \rightarrow \nu_{\mu} ; x) - 
P (\bar{\nu}_{\mu} \rightarrow \bar{\nu}_{\mu} ; x)}
{P (\nu_{\mu} \rightarrow \nu_{\mu} ; x) + 
P (\bar{\nu}_{\mu} \rightarrow \bar{\nu}_{\mu} ; x)}
\end{equation}

\noindent
and it is represented in fig. \ref{asym} as function of the energy for a
baseline of $L = 8000$ km and different values of $\sin^2{2 \, \theta_{13}}$.
Around the resonance, $A_{CPT}$ presents a plateau with non-vanishing
appreciable values (depending on $\sin^2{2 \, \theta_{13}}$). The big
asymmetries at 6 and 20 GeV correspond to low probabilities and they are not of
interest. The negative (positive) asymmetry in the plateau is obtained for
$\Delta m^{2}_{31} > 0 (< 0)$. Obviously, it is symmetric with respect to the
horizontal axis when changing the sign of $\Delta m^{2}_{31}$. As we have seen
above, the optimal baseline is inversely proportional to the $\theta_{13}$
mixing.   

%\vspace{2mm}

%\pagebreak[4]

For atmospheric neutrinos, $A_{CPT}$ cannot be separated out and the $\nu_e
(\overline{\nu}_e)$ flux also contributes to the detection of $\nu_{\mu}
(\overline{\nu}_{\mu})$. Taking into account the CC cross-sections in the
detector,

\begin{equation}
\label{nevents}
\begin{array}{c}
N(\mu^-;E) = \sigma_{cc}(\nu_{\mu}) \; \left[ 
\phi^o (\nu_{\mu};E) \; P(\nu_{\mu} \rightarrow \nu_{\mu}) +
\phi^o (\nu_e:E) \; P(\nu_e \rightarrow \nu_{\mu}) \right] \\[2ex]
N(\mu^+;E) = \sigma_{cc}(\overline{\nu}_{\mu}) \; \left[ 
\phi^o (\overline{\nu}_{\mu};E) \; P(\overline{\nu}_{\mu} \rightarrow 
\overline{\nu}_{\mu}) +
\phi^o (\overline{\nu}_e;E) \; 
P(\overline{\nu}_e \rightarrow \overline{\nu}_{\mu}) \right]
\end{array}
\end{equation}    

\noindent
where $\phi^o (\nu_{\mu};E)$ ($\phi^o (\overline{\nu}_{\mu};E)$) and $\phi^o
(\nu_e;E)$ ($\phi^o (\overline{\nu}_e;E)$) are the muon and electron
(anti) neutrino fluxes, respectively, calculated from \cite{fnv}. As in the
important energy range, both cross-sections are, to good aproximation, linear
with the energy, one can build an asymmetry which eliminates what is induced by
$\sigma_{cc}$ in the form 

\begin{equation}
\label{a}
A = \frac{N(\mu^-;E) - \frac{\sigma_{cc}(\nu_{\mu})}
{\sigma_{cc}(\overline{\nu}_{\mu})} \; N(\mu^+;E)}{N(\mu^-;E) + 
\frac{\sigma_{cc}(\nu_{\mu})}{\sigma_{cc}(\overline{\nu}_{\mu})} \; N(\mu^+;E)}
\end{equation}

In (\ref{a}) there is still some asymmetry generated by the atmospheric
neutrino fluxes. Contrary to $A_{CPT}$, the value of the muon-charge asymmetry
is not symmetric with respect to the abscisa axis when changing the sign of
$\Delta m^{2}_{31}$. In fig. \ref{asym} we give the values of A for two values
of $\sin^2{2 \, \theta_{13}}$. There is again an appreciable separation between
the cases of positive and negative $\Delta m^{2}_{31}$.

  \section{Conclusions}

In the limit of $\frac{\Delta m^{2}_{21}}{4 E} L \ll 1$, the main conclusions
of this study are: i) The medium effects, which discriminate between neutrino
and antineutrino propagation determine the sign of the atmospheric $\Delta
m^{2}_{31}$; ii) for $s_{13} = 0$, electron neutrinos decouple from neutrino
mixing in matter and have a definite effective mass in matter; iii) for $s_{13}
\neq 0$, electron neutrinos mix with the third mass eigenstate neutrino and
take part in the atmospheric neutrino oscillations; iv) electron neutrinos do
not mix with the first mass eigenstate in matter, avoiding the generation of
genuine CP-violating effects; v) non-resonant medium effects are already
apparent in the sub-sominant channel $\nu_e \rightarrow \nu_{\mu}$ for
baselines $L \sim 3000$ km, in both the mixing and the oscillation phase-shift;
vi) the observation of matter effects in the $\nu_{\mu}$-survival probability
requires the action of the MSW resonance, with baselines longer than $L \sim
7000$ km; vii) the optimal baseline depends on the value os $s_{13}$, but the
effects are much cleaner in the region of the longest baselines without
entering the Earth core \cite{earth} (nadir angles $\theta_n \gtrsim
33^o$). \\ 

\section*{Acknowledgements}
This work is supported by Grant AEN-99-0296.

\end{document}